%% file: ShinDH.tex
\documentclass[AMA,STIX1COL]{WileyNJD-v2}

\articletype{Article Type}%
\received{}
\revised{}
\accepted{}

\input{preamble.tex} 
\usepackage{xcite}
\usepackage{xr}

\makeatletter
\newcommand*{\addFileDependency}[1]{% argument=file name and extension
  \typeout{(#1)}% latexmk will find this if $recorder=0 (however, in that case, it will ignore #1 if it is a .aux or .pdf file etc and it exists! if it doesn't exist, it will appear in the list of dependents regardless)
  \@addtofilelist{#1}% if you want it to appear in \listfiles, not really necessary and latexmk doesn't use this
  \IfFileExists{#1}{}{\typeout{No file #1.}}% latexmk will find this message if #1 doesn't exist (yet)
}
\makeatother

\usepackage[flushleft]{threeparttable}

\allowdisplaybreaks

\title{Bayesian Estimation of Hierarchical Linear Models from Incomplete Data: Cluster-Level Interaction Effects and Small Sample Sizes} 

\author[1]{Dongho Shin*} 

\author[2]{Yongyun Shin}

\author[3]{Nao Hagiwara}

\authormark{Shin et al.}
\address[1,2]{\orgdiv{Department of Biostatistics}, \orgname{Virginia Commonwealth University}, \orgaddress{\state{VA}, \country{USA}}}

\address[3]{\orgdiv{Department of Public Health Sciences}, \orgname{University of Virginia}, \orgaddress{\state{VA}, \country{USA}}}

\corres{*Department of Biostatistics, Virginia Commonwealth University, One Capitol Square, 830 East Main Street., Box 980032, Richmond, Virginia 23219, USA; \email{shind10@vcu.edu}}

\abstract[Summary]{We consider Bayesian estimation of a hierarchical linear model (HLM) from partially observed data, assumed to be missing at random, and small sample sizes. A vector of continuous covariates $C$ includes cluster-level partially observed covariates with interaction effects. Due to small sample sizes from 37 patient-physician encounters repeatedly measured at four time points, maximum likelihood estimation is suboptimal. Existing Gibbs samplers impute missing values of $C$ by a Metropolis algorithm using proposal densities that have constant variances while the target posterior distributions have nonconstant variances. Therefore, these samplers may not ensure compatibility with the HLM and, as a result, may not guarantee unbiased estimation of the HLM. We introduce a compatible Gibbs sampler that imputes parameters and missing values directly from the exact posterior distributions. We apply our Gibbs sampler to the longitudinal patient-physician encounter data and compare our estimators with those from existing methods by simulation.} 


\begin{document}
\maketitle
\onehalfspacing
\section{Introduction} \label{sec:introduction}

Medical researchers frequently confront four challenges of interest in this paper. First, patients are repeatedly measured or nested within clusters, such as physicians, clinics and neighborhoods. A hierarchical linear model (HLM) is appropriate to analyze such data \citep{raudenbush2002hierarchical, goldstein2011multilevel}. Second, missing data commonly occur at the cluster level, posing a challenge to efficient and unbiased estimation of the HLM. Third, partially observed covariates may have nonlinear effects at the cluster level to make the estimation even more challenging. In particular, when two interactive cluster-level covariates are partially observed, complete-case analysis drops the clusters and all their nested units having either of the covariates missing. The resulting inference is inefficient and may be substantially biased \citep{little2002bayes}. Finally, small sample sizes constrain estimability of the HLM.

Hierarchical missing data are routinely handled by multiple imputation (MI) \citep{rubin1987} under the assumption of data missing at random (MAR) \citep{rubin1976inference}. A popular joint modeling approach conceives of the response and partially observed covariates as the outcomes in a multivariate HLM, estimates their joint distribution, and imputes missing values from the estimated distribution \citep{schafer2002computational, shin2007just}. Missing data may also be imputed from their conditional distributions by chained regressions, which is also known as Fully Conditional Specification (FCS) \citep{raghunathan2001multivariate, van2007multiple, van2011mice, van2018flexible}. FCS fits each partially observed variable conditional on all other variables and imputes missing values from the fitted model. FCS is a Gibbs sampler if the chained regressions are compatible with an underlying joint distribution of all variables MAR, including the outcome \citep{arnold1989compatible, arnold2001conditionally, liu2014stationary, bartlett2015multiple}. These approaches have been shown to work well when the joint distribution of partially observed variables are reasonably multivariate normal \citep{liu2014stationary, bartlett2015multiple, enders2016multilevel}.

Given the nonlinear effects of covariates $C$ on a response $Y$ in a HLM, however, multivariate normality of $(Y,C)$ is not possible even if the factorized conditional distributions are normal \citep{kim2015evaluating, shin2024maximum}. A concern then involves the bias that can arise from the incompatibility or conflicting assumptions between the fully conditional distributions that generate the imputations and the assumed joint distribution of the observed data \citep{carpenterandkenward, liu2014stationary, bartlett2015multiple, kim2015evaluating, erler2016dealing, enders2020model}. Kim et al.\citep{kim2015evaluating, kim2018multiple} developed a Gibbs sampler to impute missing values of a partially observed continuous covariate interactive with a known covariate in a single-level regression model from the exact conditional distribution. Carpenter and Kenward\citep{carpenterandkenward} took a multivariate normal (MN) approach to handling continuous and categorical data MAR by joint normality of continuous variables and latent continuous variables underlying categorical ones. Goldstein et al.\citep{goldstein2014fitting} imputed missing values of covariates having nonlinear effects on a continuous or binary outcome by the MN approach via the Gibbs sampler and Metropolis algorithm \citep{metropolis1953equation,hastings1970monte}. Enders et al.\citep{enders2020model} extended the MN-Metropolis Gibbs sampler to handling categorical and continuous predictors MAR having nonlinear effects in two- and three-level hierarchical models, and illustrated how FCS using its software MICE \citep{van2011mice} may lead to an imputation model incompatible with the HLM to produce biased estimates.

Shin and Raudenbush\citep{shin2024maximum} extended the joint modeling approach to estimation of a HLM with the non-linearities from continuous data $(Y,C)$ MAR. The HLM implied a compatible nonstandard joint distribution $h(Y,C)$ such that the likelihood function $L(\theta)$ of the joint parameters $\theta$ is an integral with respect to high-dimensional random effects $(u,\nu)$ given the observed data $(Y_{obs},C_{obs})$. They selected ``provisionally known random effects (PKREs)'' $u$ such that $h(Y_{obs},C_{obs}|u)=\int h(Y_{obs},C_{obs},\nu|u)d\nu$ is analytically tractable with respect to high dimensional $\nu$ and, then, $L(\theta)=\int h(Y_{obs},C_{obs}|u)g(u)du$ is approximated numerically with respect to low dimensional $u$ by adaptive Gauss-Hermite quadrature (AGHQ). They maximized the likelihood  by the EM algorithm to produce the estimates $\hat{\theta}$ at ML, and translated the $\hat{\theta}$ to the ML estimates of the HLM. They illustrated efficient ML estimation of random and fixed interaction effects involving lower-level covariates and their latent cluster means. Given the estimates $\hat{\theta}$ and associated $var(\hat{\theta})$ at ML, Shin and Raudenbush\citep{shin2024b} generated MI of missing data, including random effects, from a nonstandard predictive model $h(Y, C, \nu, u |  Y_{obs}, C_{obs}, \theta)$ by importance sampling via AGHQ, and estimated the HLM given the MI. These approaches within the Maximum Likelihood framework, however, are suboptimal given small sample sizes.

We now explain our contribution to the literature. Unlike existing samplers that sample missing values of a cluster-level covariate having a nonlinear effect by a metropolis algorithm via a proposal normal density with a constant variance \citep{goldstein2014fitting, enders2020model}, our Gibbs sampler samples missing data and parameters from their exact posterior distributions. Therefore, our Gibbs sampler is guaranteed to be compatible and, thus, to produce unbiased estimation of an analytic HLM with non-linear effects of cluster-level covariates. Either one or both of the interactive covariates may be partially observed and partially observed covariates are continuous. The results of the simulation study below reveal that our Gibbs sampler is particularly effective with comparatively small sample sizes.

Our motivating application is the analysis of racially discordant patient-physician medical interactions, focusing on pairs of patients and physicians from different racial groups during office visits. These visits were videotaped and coded to produce four repeated measurements of the outcome: a positive valence score measuring the physician's facial expression. Of interest are the main and interaction effects of continuous physician's implicit and explicit prejudices on the outcome that may be appropriately analyzed using a HLM. Three challenges motivated us to develop our Gibbs sampler. First, valence score is missing 20\% of the values, and each of the two key physician covariates has 16\% missing data at the cluster level. Next, the effects of partially observed physician prejudices include an interaction effect. Finally, due to the COVID restrictions, only 37 encounters were recorded from the small number of 6 physicians and 37 patients, posing a formidable challenge to the efficient and unbiased estimation of the HLM.

The rest of the paper is organized as follows. Section 2 introduces our general HLM with the nonlinear effects of cluster-level covariates. Section 3 presents our Gibbs sampler. Section 4 evaluates our sampler by comparing our estimators with those by existing methods in simulated analyses. Section 5 illustrates analysis of real data from the racially discordant patient-physician encounters using our Gibbs sampler. Lastly, section 6 discusses limitations and future extensions of our approach.

%%%%%%%%%%%%%%%%%%%%%%%%%%%%%%%%%%%%%%%%%%
\section{Model}
Our interest focuses on a two-level hierarchical linear model (HLM) 
\begin{flalign} 
\label{HLM1}
Y_{ij}=\beta_{0}+\beta^T_CC_{j}+\beta^T_XX_{ij}+\sum_{s=1}^{p}\beta^T_{XCs}X_{ij}C_{sj}+\sum_{s=1}^{p-1}\sum_{t=s+1}^{p}\beta_{CCst}C_{sj}C_{tj}+u_{j}+e_{ij}, 
\end{flalign}
\noindent where $Y_{ij}$ is the outcome variable, $C_j=[C_{1j} ... C_{pj}]^T$ is a $p$-by-$1$ vector of partially observed cluster-level continuous covariates having main effects $\beta_C$, and $X_{ij}=[x^T_{1ij}\; x^T_{2j}]$ is a $q$-by-$1$ vector of fully observed covariates having main effects $\beta_X=[\beta^T_{x1}\; \beta^T_{x2}]^T$ for lower-level or level-1 covariates $x_{1ij}$ and cluster-level or level-2 covariates $x_{2j}$. In addition, $\beta_{XCs}$ is the q-by-1 interaction effects of $X_{ij}$ and $C_{sj}$, while $\beta_{CCst}$ represents the scalar interaction effect of $C_{sj}$ and $C_{tj}$ for $s \le t$. Lastly, a level-2 unit-specific random effect $u_j\sim \mathcal{N}(0,\tau)$ and a level-1 unit specific random effect $e_{ij} \sim N(0,\sigma^2)$ are independent, and a level-1 unit $i$ is nested within a level-2 cluster $j$ for $i=1,\cdots,n_j$ and $j=1,\cdots,J$. Here, the covariates $C_{j}$ and the outcome $Y_{ij}$ may be partially observed.

%%%%%%%%%%%%%%%%%%%%%%%%%%%%%%%%%%%%%%%%%%
\section{Compatible Gibbs Sampler}
This section explains our Gibbs sampler based on the exact posteriors of parameters and missing data, unlike existing Bayesian approaches using rejection sampling of missing data by the Metropolis algorithm \citep{goldstein2014fitting, enders2020model}.  To handle missing data efficiently, we assume the joint normal distribution of $C_{j}$ conditional on known covariates to write our Bayesian joint distribution of $\mathbf{Y}=(Y_{11},Y_{12},\cdots,Y_{n_JJ}), 
 \mathbf{C}=(C_1,C_2,\cdots,C_J)$ and $\theta$ given $\mathbf{X}=(X_{11},X_{12},\cdots,X_{n_JJ})$
\begin{align}
\label{joint}
    f(\mathbf{Y},\mathbf{C},\theta|\mathbf{X})= 
    \prod_{j=1}^J \prod_{i=1}^{n_j}f(Y_{ij}|C_{j},X_{ij},u_j,\boldsymbol{\beta}, \sigma^2)f(u_j|\tau)f(C_{j}|x_{2j},\alpha,T)p(\theta)
\end{align}
\noindent where $p(\theta)$ is the prior distribution of $\theta=(\boldsymbol{\beta}, \tau, \sigma^2, \alpha,T)$, as specified in the Gibbs sampler steps below, with $\boldsymbol{\beta}=(\beta_0, \beta_C, \beta_X, \beta_{XC}, \beta_{CC})$ for $\beta_{XC}=(\beta_{XC1},\beta_{XC2},...,\beta_{XCp})$ and $\beta_{CC}=(\beta_{CC12},...,\beta_{CC1p}, \beta_{CC23}...,\beta_{CC2p},...,\beta_{(p-1)p}).$ Here $f(Y_{ij}|C_{j},X_{ij},u_j,\boldsymbol{\beta},\sigma^2)$ and $f(u_j|\tau)$ are normal densities from the HLM (\ref{HLM1}), and
\begin{align}
\label{covariate model}
f(C_{j}|x_{2j}) \sim N\left[ W\alpha = \left(I_p \otimes \left[1\; x^T_{2j}\right]\right) \alpha, T\right]
\end{align}
\noindent for a vector $\alpha$ of fixed effects, a $p\times p$ identity matrix $I_p$, a kronecker product $A\otimes B$ multiplying matrix B to each element of matrix A, and a $p\times p$ variance-covariance matrix $T$. To derive the Gibbs sampler, we partition complete data $\mathbf{Y}=(\mathbf{Y}_{obs},\mathbf{Y}_{mis})$ and $\mathbf{C}=(\mathbf{C}_{obs},\mathbf{C}_{mis})$ into observed  $(\mathbf{Y}_{obs}, \mathbf{C}_{obs})$ and missing $(\mathbf{Y}_{mis},\mathbf{C}_{mis})$.

\subsection{Exact Posterior Distributions of $C_{mis}$}
Let $p(A|\cdot)$ denote the posterior, or exact posterior, distribution of A given all other unknowns. To find the key posterior distribution $p(C_{kj}|\cdot)$ for a missing element $C_{kj}$ of $C_j$, we first derive a bivariate normal conditional distribution from Equations (\ref{HLM1}) and (\ref{covariate model})
\begin{gather}
\label{yc_joint}
 \left[\begin{array}{c} Y_{ij}\\ C_{kj} \end{array}\right] \left|C_{(-k)j},X_{ij},u_j,\boldsymbol{\beta}, \sigma^2,\alpha,T \right. \sim N\left(\begin{bmatrix}
 \mu_{1ij}+\mu_{2ij}M_{k|(-k)}\\ M_{k|(-k)}
\end{bmatrix}, \begin{bmatrix}
\mu_{2ij}^2T_{k|(-k)}+\sigma^2 & \mu_{2ij}T_{k|(-k)}  \\
\mu_{2ij}T_{k|(-k)} & T_{k|(-k)}  \\
    \end{bmatrix}\right)
    \intertext{for $C_{(-k)j}=(C_{1j},...,C_{(k-1)j},C_{(k+1)j},...,C_{pj})$, $M_{k|(-k)}=E(C_{kj} | C_{(-k)j})$ and $T_{k|(-k)}=var(C_{kj} | C_{(-k)j})$. Shin and Raudenbush \citep{shin2024maximum} referred to missing values of $C_{(-k)j}$ as provisionally known random effects. The conditional mean $E(Y_{ij}|C_{(-k)j}, X_{ij}, u_j)$ has two parts, $\mu_{1ij}$ excluding and $\mu_{2ij}$ including $C_{kj}$, to facilitate derivation of $p(C_{kj}|\cdot)$:}
    \begin{aligned}
&\mu_{1ij}=\beta_{0}+\beta^T_{C(-k)}C_{(-k)j}+\beta^T_XX_{ij}+\sum_{s\ne k, s \ge 1}^{p}\beta^T_{XCs}X_{ij}C_{sj}+\sum_{s\ne k, s\ge 1}^{p-1}\sum_{t \neq k, t \ge s}^{p}\beta_{CCst}C_{sj}C_{tj}+u_j, \\ \nonumber
&\mu_{2ij}
=\left(\beta_{C_k}+\beta^T_{XC_k}X_{ij}+\sum_{s=1}^{p-1}\sum_{t=s+1}^{p}\left(\beta_{CCst}C_{tj}I(s=k)+\beta_{CCst}C_{sj}I(t=k)\right)\right)
\end{aligned}
\end{gather}
for an indicator function $I(B)=1$ if condition $B$ is true and 0 otherwise, $\beta_C$ partitioned into the coefficients $\beta_{C_k}$ of $C_{kj}$ and $\beta_{C(-k)}$ of $C_{(-k)j}$, and $\sum_{s\neq k, s\geq a}^{p} A_s = \sum_{s=a}^{k-1} A_s + \sum_{s=k+1}^{p} A_s.$

The joint distribution (\ref{joint}) then implies
\begin{align}
\label{imputation_model}
p(C_{kj}|\cdot) & \propto \prod_{i=1}^{n_j}f(Y_{ij}|C_{j},X_{ij},u_j,\boldsymbol{\beta}, \sigma^2) \times f(C_{kj}|C_{(-k)j},x_{2j}, \alpha, T) \sim N(\Tilde{M}_{kj}, \Delta_{kj}^{-1}),
\end{align}
where $\Tilde{M}_{kj}=M_{k|(-k)}+\Delta^{-1}_{kj}\sigma^{-2}\mu_{2ij}\sum_{i=1}^{n_j}\left(Y_{ij}-(\mu_{1ij}+\mu_{2ij}M_{k|(-k)})\right)$ and $\Delta_{kj}=T^{-1}_{k|(-k)}+n_j\mu_{2ij}^2\sigma^{-2}.$

\subsection{Gibbs Sampler Steps Based on  \textcolor{black}{Exact Posterior Distributions}}
The joint distribution (\ref{joint}) implies the posterior distributions of $\mathbf{u}=(u_1,\cdots,u_J)$ and $\theta$:
\begin{align}
 &p(u_j|\cdot) \propto \prod_{i=1}^{n_j}f(Y_{ij}|C_{j}, X_{ij}, u_j, \boldsymbol{\beta}, \sigma^2)f(u_j|\tau), \\
&p(\tau|\cdot) \propto \prod_{j=1}^{J}f(u_j|\tau)p(\tau), \nonumber \\
&p(\boldsymbol{\beta}|\cdot) \propto \prod_{j=1}^{J}\prod_{i=1}^{n_j}f(Y_{ij}|C_{j}, X_{ij},u_j, \boldsymbol{\beta}, \sigma^2)f(\boldsymbol{\beta}), \nonumber \\
&p(\sigma^2|\cdot) \propto \prod_{j=1}^{J}\prod_{i=1}^{n_j}f(Y_{ij}|C_{j},X_{ij},u_j, \boldsymbol{\beta}, \sigma^2)p(\sigma^2), \nonumber \\
&p(\alpha|\cdot) \propto \prod_{j=1}^{J}f(C_{j}|x_{2j},\alpha,T)p(\alpha), \nonumber \\
&p(T|\cdot) \propto \prod_{j=1}^{J}f(C_{j}|x_{2j},\alpha,T)p(T),  \nonumber 
\end{align}

\noindent where we assume inverse gamma priors $p(\tau)\sim IG(\alpha_0=1, \beta_0=0.5)$ and $p(\sigma^2)\sim IG(\alpha_0=1, \beta_0=0.5)$, an inverse wishart $p(T)\sim IW(V_0,S_0^{-1})$ with $V_0=p+2$ and $S_0=\hat{T}$ (an estimated variance-covariance matrix using complete cases), and non-informative priors $p(\beta)=p(\alpha)=1$, following Schafer and Yucel\citep{schafer2002computational}, Hoff\citep{hoff2009first} and Enders et al.\citep{enders2020model}.

At cycle $t$, we update the current values of parameters $\theta=\theta^{(t-1)}$ and completed data $\mathbf{Y}=(\mathbf{Y}_{obs},\mathbf{Y}^{(t-1)}_{mis})$ and $\mathbf{C}=(\mathbf{C}_{obs},\mathbf{C}^{(t-1)}_{mis})$ from cycle $t-1$ in eight steps, as follows:

\begin{enumerate}[label=Step \arabic*:,leftmargin=*]
\item Sample $u^{(t)}_j$ from 
\begin{align*}
  p(u_j|\mathbf{Y}=\mathbf{Y}^{(t-1)},\theta=\theta^{(t-1)}) \sim N(\Delta_j^{-1}\sigma^{-2} \sum_{i=1}^{n_j}(Y_{ij}-\mathbf{X}^T_{ij}\boldsymbol{\beta}), \Delta_j^{-1}),
\end{align*}     
\noindent where $\mathbf{X}^T_{ij}=\left[1\; C^T_{j}\; X^T_{ij}\; X^T_{ij}\otimes C^T_{j}\; \text{vech}^T\left(C_{(-1)j}C^T_{(-p)j}\right)\right]$, $\boldsymbol{\beta}=\left[\beta_0\; \beta^T_C\; \beta^T_X\; \beta^T_{XC}\; \beta^T_{CC}\right]^T$ and $\Delta_j=n_j\sigma^{-2}+\tau^{-1}$. Here, the operator $\text{vech}(\cdot):\mathbb{R}^{r\times r}\rightarrow \mathbb{R}^{r(r+1)/2}$ stacks the upper triangular entries of an $r \times r$ square matrix into a column vector. 

\item Draw $\tau^{(t)}$ from
\begin{align*}
    p(\tau | \mathbf{u}=\mathbf{u}^{(t)}, \theta=\theta^{(t-1)}) \sim IG\left(\frac{J}{2}+\alpha_0, \Bigr[\frac{\sum_{i=1}^{n_j}u_j^2}{2}+\frac{1}{\beta_0}\Bigr]^{-1}\right).
\end{align*} \par

\item Draw $\boldsymbol{\beta}^{(t)}$ from
\begin{align*}
    p(\boldsymbol{\beta}|\mathbf{Y}=\mathbf{Y}^{(t-1)}, \mathbf{u}=\mathbf{u}^{(t)}, \sigma^{2 (t-1)}) \sim N\left(\Bigr(\sum_{j=1}^{J}\sum_{i=1}^{n_j}\mathbf{X}_{ij}\mathbf{X}_{ij}^T\Bigr)^{-1}\sum_{j=1}^{J}\sum_{i=1}^{n_j}\mathbf{X}_{ij}(Y_{ij}-u_j),\sigma^2\Bigr(\sum_{j=1}^{J}\sum_{i=1}^{n_j}\mathbf{X}_{ij}\mathbf{X}_{ij}^T\Bigr)^{-1}\right). 
\end{align*} 

\item Draw $\sigma^{2(t)}$ from 
\begin{align*}
    p(\sigma^2|\mathbf{Y}=\mathbf{Y}^{(t-1)}, \mathbf{u}=\mathbf{u}^{(t)}, \boldsymbol{\beta}=\boldsymbol{\beta}^{(t)}) \sim IG\left(\frac{N}{2}+\alpha_0, \Bigr[\frac{\sum_{j=1}^{J}\sum_{i=1}^{n_j}e_{ij}^2}{2}+\frac{1}{\beta_0}\Bigr]^{-1}\right),
\end{align*} 
where $N=\sum_{j=1}^J n_j$.

\item For $Y_{ij}$ missing, impute $e^{(t)}_{ij}$ from $p(e_{ij}|\sigma^2=\sigma^{2 (t)}) \sim N(0,\sigma^2)$ and set $Y^{(t)}_{ij}=\mathbf{X}^T_j\boldsymbol{\beta}^{(t)}+u^{(t)}_j+e^{(t)}_{ij}.$ 

\item Draw $\alpha^{(t)}$ from
\begin{align*}
    p(\alpha|\mathbf{C}=\mathbf{C}^{(t-1)}, T=T^{(t-1)}) \sim N\left(\Bigr(\sum_{j=1}^{J}W^TT^{-1}W\Bigr)^{-1}\sum_{j=1}^{J}W^TT^{-1}C_j, \Bigr(\sum_{j=1}^{J}W^TT^{-1}W\Bigr)^{-1}\right).
\end{align*} 

\item Draw $T^{(t)}$ from
\begin{align*}
    p(T | \mathbf{C}=\mathbf{C}^{(t-1)}, \alpha = \alpha^{(t)}) \sim IW\left(V_0+J, \Bigr(S_0+\sum_{j=1}^{J}(C_j-W\alpha)(C_j-W\alpha)^T\Bigr)^{-1}\right). 
\end{align*} 

\item Repeat the following sub-step as many times as the number of missing values in $C_j$. For each $C_{kj}$ missing, define
\begin{align*}
    C^{(t)}_{(-k)j}=(C^{(t)}_1,...,C^{(t)}_{k-1}, C^{(t-1)}_{k+1},...,C^{(t-1)}_p) 
\end{align*}
that consists of $(k-1)$ observed or imputed missing values at cycle $t$ and $(p-k)$ observed or imputed values at cycle $t-1$. We compose a bivariate distribution (\ref{yc_joint}):
\begin{align*}
    f(Y_{ij},C_{kj}|\mathbf{Y}=\mathbf{Y}^{(t)}, C_{(-k)j}=C^{(t)}_{(-k)j}, u_j=u^{(t)}_j,\theta=\theta^{(t)})
\end{align*}
to draw $C_{kj}$ from the implied posterior distribution (\ref{imputation_model}):
\begin{align*}
    p(C_{kj}| \mathbf{Y}=\mathbf{Y}^{(t)}, C_{(-k)j}=C^{(t)}_{(-k)j}, u_j=u^{(t)}_j,\theta=\theta^{(t)}).
\end{align*}
\end{enumerate}
\clearpage
\section{Simulation Study}

We now assess our compatible Gibbs sampler based on exact posterior distributions (GSExact) using simulated data under \textcolor{black}{four different scenarios: i) correctly specified distributions of the response, covariates and missing data mechanism; ii) misspecified distribution of a covariate; iii) violation of the MAR assumption; and iv) controlling for multiple partially observed interaction terms.} The latter three cases evaluate GSExact in terms of robust estimation against a misspecified covariate distribution, a violated missing data mechanism, and additional partially observed nonlinear terms, respectively. The simulated data from HLM (\ref{HLM1}) will closely resemble real data that we analyze in the next section in terms of correlations, sample sizes and missing rates. In each case, we compare our estimators with those from: 1) the lme4 package in R \citep{douglas2015fitting} that estimates maximum likelihood estimates given complete data (CDML) and 2) software Blimp \citep{blimp} that implements the Gibbs Sampler of Enders et al.\citep{enders2020model}. We do not consider FCS, as it has been shown to be incompatible and produce biased estimates due to nonlinear effects of covariates MAR\citep{liu2014stationary, bartlett2015multiple, enders2016multilevel, enders2020model}. However, we compare GSEsxact with imputations by predictive mean matching in terms of robust estimation in the Appendix. Because CDML estimators are estimated from complete data while others are based on data MAR, a good method will produce estimates near CDML counterparts. 
\textcolor{black}{\subsection{Correctly Specified Distributions of the Response, Covariates and MAR Mechanism}}
We consider two cases where $n_j=4$ units are nested within each of 1) $J=36$ clusters or a small sample and 2) $J=200$ clusters or a large sample. We validate the correct execution of the R code that implements GSExact given the large sample size and and compare our estimators with the competing ones given the small and large sample sizes. 

We simulate sequentially: i) $X_j\sim N(2,1)$; ii) $C_{1j}\sim N(0.75+0.7X_j, 1.25)$ and $ C_{2j} \sim N(-0.5+X_j, 1)$ with $T_{12}=cov(C_{1j},C_{2j})=-0.5$ in model (\ref{covariate model}); and iii) $Y_{ij}\sim N(1+C_{1j}+C_{2j}+X_j+C_{1j}C_{2j}, \tau+\sigma^2)$ for $\tau=4$ and $\sigma^2=16$ in HLM (\ref{HLM1}). The simulated coefficients of the HLM are all equal to 1 to facilitate the comparison.

Next, we simulate the missing values of $Y_{ij}$, $C_{1j}$ and $C_{2j}$ by a MAR mechanism that depends on the fully known $X_j$ from $Bernoulli(p_j)$ where
\begin{align}
\label{missing_mechanism}
logit(p_j) \sim N(c_0+c_1X_j, \delta). 
\end{align}
\noindent We manipulate $c_0$, $c_1$ and $\delta$ to simulate 20\% missing values for each of the following variables: $Y_{ij}$ given $c_0=-1.9, c_1=0.1$ and $\delta=1$, $C_{1j}$ given $c_0=0.8, c_1=-1.5$ and $\delta=0$, and $C_{2j}$ given $c_0=-2.8, c_1=0.5$ and $\delta=0$.

\textcolor{black}{We repeated simulating data and estimating the simulated HLM 1,000 times to compute the \% bias equal to (estimated - simulated) * 100 / simulated,} average estimated standard error (ASE), empirical estimate of the true standard error (ESE) over the samples and the coverage probability (coverage) of each estimator. Both Blimp and GSExact are based on 2,500 burn-in and 2,500 post burn-in iterations. Given each simulated data set, we run two separate Gibbs sampler chains given different initial values to evaluate the convergence by Geweke's method \citep{geweke1991evaluating} and the potential scale reduction factor (PSRF) \citep{brooks1998general}. 

Table \ref{j=200} summarizes the simulation results for the large sample size. Both GSExact and Blimp produce estimates close to those by CDML. All three approaches produce very accurate and precise estimates with biases $<$ 2\% except those of the $\tau$ estimates by GSExact (-2.5\%) and Blimp (4.9\%), ASEs close to ESEs, and coverage probabilities near the nominal 0.95. The standard errors by GSExact and Blimp are comparatively inflated to reflect extra uncertainty from missing data.

\begin{table}[htbp]
\centering
  \caption{Estimated \% biases, ASEs, ESEs and coverages from 1,000 simulated large-sample data sets $(n_j=4, J=200)$.}
\begin{adjustbox}{max width=\textwidth}
\renewcommand{\arraystretch}{1.5}
    \begin{tabular}{cccccccccc}
\toprule
  & \multicolumn{3}{c}{CDML} & \multicolumn{3}{c}{GSExact} & \multicolumn{3}{c}{Blimp} \\
\cmidrule(lr){2-4}\cmidrule(lr){5-7}\cmidrule(lr){8-10}
 Simulated & \%Bias (ASE) & ESE & Coverage & \%Bias (ASE) & ESE & Coverage  & \%Bias (ASE) & ESE & Coverage \\ \midrule
 
 $\tau$=4 & 0.6 (0.84) & 0.85 & 0.95 & -2.5 (1.07) & 1.05  & 0.95 & 4.9 (1.13) & 1.07 & 0.95 \\

 $\sigma^2$=16 & 0.1 (0.92) & 0.88 & 0.95 & 0.8 (1.08) & 1.03 & 0.96 & 0.5 (1.08) & 1.02  & 0.96  \\

 $\beta_{0}$=1 & -1.7 (0.58) & 0.58 & 0.95 &0.2 (0.66) & 0.65 & 0.95 & 0.1 (0.67) & 0.65 & 0.95 \\

 $\beta_{1}$=1 & -0.6 (0.26) & 0.26 & 0.95 & 0.0 (0.32) & 0.32 & 0.94 & -0.3 (0.32) & 0.32 & 0.96 \\

 $\beta_{2}$=1 & 1.0 (0.33) & 0.34 & 0.94 & -0.3 (0.39) & 0.41 & 0.94 & -0.7 (0.40) & 0.41 & 0.94 \\

 $\beta_{3}$=1 & 0.5 (0.37)  & 0.38 & 0.94 & 0.5 (0.45) & 0.46  & 0.94 & 1.6 (0.46) & 0.46 & 0.94 \\

 $\beta_{4}$=1 & -0.5 (0.11)  & 0.11 & 0.95 & -0.1 (0.13) & 0.13  & 0.94 & -0.3 (0.14) & 0.14 & 0.95\\
\bottomrule
\end{tabular}%
\end{adjustbox}
\label{j=200}%
\end{table}%

\begin{table}[htbp]
\centering
  \caption{\textcolor{black}{Estimated \% biases, ASEs, ESEs and coverages from 5,000 simulated small-sample data sets $(n_j=4, J=36)$.}}
\begin{adjustbox}{max width=\textwidth}
\renewcommand{\arraystretch}{1.5}
    \begin{tabular}{cccccccccc}
\toprule
  & \multicolumn{3}{c}{CDML} & \multicolumn{3}{c}{GSExact} & \multicolumn{3}{c}{Blimp} \\
\cmidrule(lr){2-4}\cmidrule(lr){5-7}\cmidrule(lr){8-10}
 Simulated & \%Bias (ASE) & ESE & Coverage & \%Bias (ASE) & ESE & Coverage  & \%Bias (ASE) & ESE & Coverage \\ \midrule
 
 $\tau$=4 & -0.2 (2.10) & 2.07 & 0.96 & -1.9 (2.54) & 2.00  & 0.95 & 25.9 (3.57) & 2.60 & 0.97 \\

 $\sigma^2$=16 & 0.2 (2.18) & 2.20 & 0.95 & 1.9 (2.57) & 2.49 & 0.94 & 2.5 (2.66) & 2.51 & 0.95  \\

 $\beta_{0}$=1 & -3.5 (1.51) & 1.55 & 0.95 & 6.3 (1.89) & 1.84 & 0.97 & 4.1 (2.16) & 1.90 & 0.97 \\

 $\beta_{1}$=1 & -0.8 (0.69) & 0.69 & 0.94 & -0.9 (0.90) & 0.88 & 0.97 & -4.7 (1.02) & 0.90 & 0.96 \\

 $\beta_{2}$=1 & -0.3 (0.86) & 0.87 & 0.95 & -0.9 (1.13) & 1.13 & 0.96 & -5.9 (1.26) & 1.11 & 0.97 \\

 $\beta_{3}$=1 & -1.1 (0.94)  & 0.94 & 0.94 & -3.8 (1.19) & 1.20  & 0.97 & 8.9 (1.31) & 1.19  & 0.96 \\

 $\beta_{4}$=1 & 0.3 (0.30)  & 0.31 & 0.94 & 1.3 (0.40) & 0.39  & 0.96 & 0.3 (0.44) & 0.39  & 0.97\\
\bottomrule
\end{tabular}%
\end{adjustbox}
\label{j=36}%
\end{table}%

Table \ref{j=36} summarizes the results from the small sample simulation. Because the small sample size led to greater sampling variability compared to the large sample size, we ran 5,000 simulations to reduce this variability. The CDML estimates are consistently close to simulated values with bias $<$ -3.5\% in magnitude and small ASEs close to ESEs, achieving good coverages near the nominal 0.95. GSExact estimates are reasonably close to the CDML estimates with small biases $<$ 2\% in magnitude except two fixed effects. The estimator for the intercept $\beta_0$ has a comparatively large bias of 6.3\%, which is 2.8\% larger in magnitude than the -3.5\% bias of the CDML counterpart. Additionally, the effect of the known covariate $X_j$, on which the missing pattern depends, showed a -3.8\% bias in the estimation. ASEs are small near ESEs with good coverages. 

Blimp produces comparatively large biases overall. In particular, the biases in the main effects ($\beta_{1}$, $\beta_{2}$ and $\beta_{3}$) are noticeably larger than those of GSExact, but the estimated interaction effect $\beta_4$ is very accurate. The ASEs are near ESEs, and overall modestly inflated compared to those of GSExact. As a result, the coverages are good near the nominal level. The level-2 variance estimate of $\tau$ is biased upward by 25.9\%, the largest bias observed, which is larger in magnitude than the GSExact counterpart bias of -1.9\%. Additionally, the ASE of the $\tau$ estimate is 3.57, which is 41\% higher than the GSexact counterpart of 2.54. Therefore, small sample estimates differ noticeably between GSExact based on exact posterior distributions and Blimp based on a metropolis algorithm in sampling missing values of $C_{1j}$ and $C_{2j}$ from normal proposal densities with constant variances. Overall, GSExact and Blimp ASEs are again larger than the CDML counterparts, which reflects extra uncertainty due to missing data.

\textbf{Convergence.} We assessed convergence to a stationary distribution by both Geweke's method \citep{geweke1991evaluating} and potential scale reduction factor (PSRF) \citep{brooks1998general}. Geweke's statistic tests if the two means of an estimator, typically from the first 20\% and the last 50\% of post burn-in iterations, are equal within a single Markov chain by the two-sample normal Z test. On the contrary, the PSRF compares the within-chain and between-chain variances from multiple Markov chains--two in our case--to evaluate convergence. A PSRF value $<$ 1.1 indicates convergence \citep{du2022performances, cowles1996markov}. We assess the convergence of GSExact as all 7 estimators of the HLM satisfying each convergence criterion, and compare the results from both criteria.

Our simulation revealed that for the large sample, only 64.5\% out of the 1,000 simulations satisfied the Geweke's convergence criterion, whereas 99\% met the PSRF criterion. Likewise, for the small sample, 66.7\% satisfied the Geweke's criterion while 99\% met the PSRF criterion. The converged and non-convergent estimates based on Geweke's criterion revealed virtually no differences in terms of \% biases, ASEs, ESEs and coverages. 

The Geweke's statistics seem to inflate the Type I error probability in multiple comparisons of the seven convergences. The rejection rates are quite close to the overall Type I error probability under the unrealistic assumption of independent estimators $1-0.95^7=0.301$, implying that it is easier to reject convergence by the Geweke's statistics under the null hypothesis of convergence. Furthermore, the Geweke's statistics rely on a single MCMC chain and, thus, is based on the within-chain variance only. Consequently, given small sample sizes where ASEs are expected to be comparatively high, we particularly expect the non-convergence rates by Geweke's criterion to be also high. Therefore, we used the PSRF criterion to assess the convergence in real analysis below. 

\begin{figure}[ht]
\centering
\begin{minipage}[b]{0.48\textwidth}
\centering
\includegraphics[width=\textwidth]{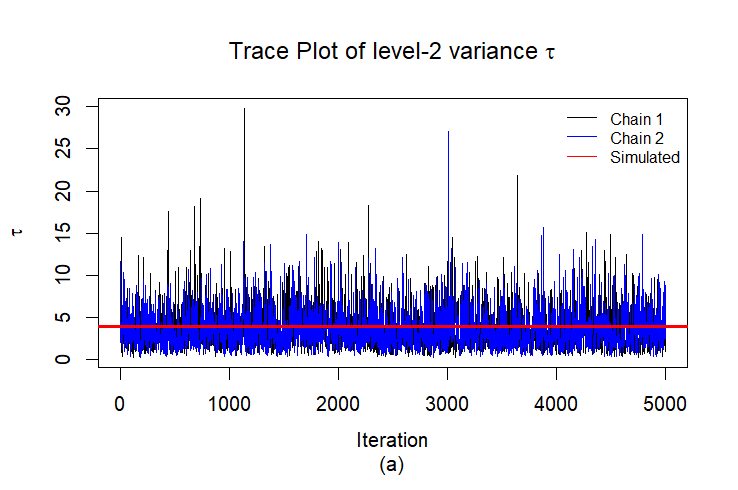}
\end{minipage}
\hfill
\begin{minipage}[b]{0.48\textwidth}
\centering
\includegraphics[width=\textwidth]{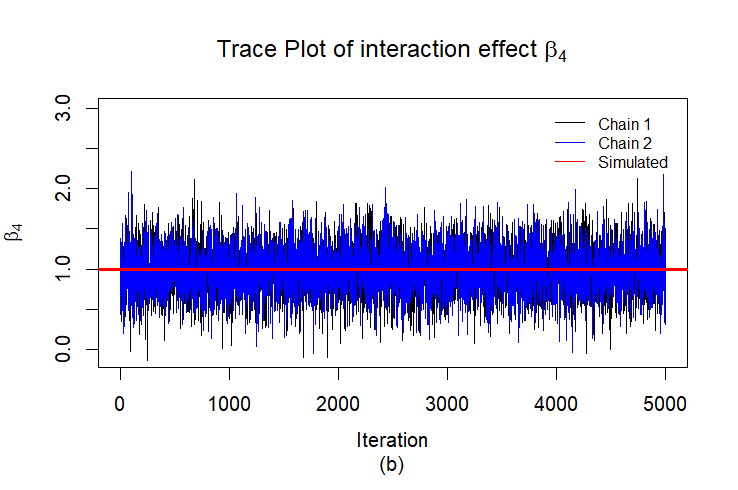}
\end{minipage}
\caption{\textcolor{black}{(a) Trace plot for the level-2 variance, $\tau$; (b) Trace plot for the interaction effect, $\beta_4$}}
\label{fig:traceplots}
\end{figure}

\textcolor{black}{Additionally, we assessed convergence using trace plots from two chains with different initial values.} Chain 1 initializes the missing data by predictive mean matching, using the mice R package, whereas chain 2 imputes them by sample means. Next, both chains fit the HLM (\ref{HLM1}) to initialize the parameters. Each trace plot tracks the sampled values of a parameter from its posterior on the vertical axis, with iteration 1 through 5,000 on the horizontal axis. Figure \ref{fig:traceplots} shows the trace plots for the level-2 variance $\tau$ and the interaction effect $\beta_4$ from one of the simulated small samples. See the Appendix for additional trace plots. Each plot exhibits evidence of convergence to a stable distribution with rapid random fluctuations around a stable horizontal band, no trends, and extensive overlap between chains, indicating exploration of the same posterior distribution.

\textcolor{black}{Although the trace plot for $\tau$ in Figure \ref{fig:traceplots}(a) oscillates randomly around its mean}, it also exhibits some large spikes to reveal considerable uncertainly in the estimation of $\tau$ from the small number of 36 clusters and high missing rates. Additionally, these spikes reflect the fact that the posterior inverse gamma distribution of $\tau$, which is right-skewed and heavy tailed, is more likely to produce large sampled values compared to the symmetric normal posterior distributions of fixed effects.

\vspace*{-1cm} % Adjust this value to reduce the gap
\textcolor{black}{\subsection{A Misspecified Covariate Distribution}}
To investigate the robustness of our estimators against violations of the normality assumption for covariates, we simulate: 1) $X_j\sim N(2,1)$; 2) $C_{1j}\sim logNormal(0.5+0.1X_j, 0.2),  C_{2j} \sim N(1+0.1C_{1j}+0.3X_j,1)$ and 3) $Y_{ij}\sim N(1+C_{1j}+C_{2j}+X_j+C_{1j}C_{2j}, \tau+\sigma^2)$ for $\tau=4$ and $\sigma^2=16$. The parameters of the log-normal distribution are set to generate $C_{1}$ with a mean of around 2 and variance near 1, and a skewness of 1.6. Therefore, the bivariate normality assumption of covariates in both GSExact and Blimp is violated. We again simulate 1,000 large and 5,000 small data sets, using the same missing rates as before.

\begin{table}[htbp]
\centering
  \caption{Estimated \% biases, ASEs, ESEs and coverages from 1,000 simulated large data sets $(n_j=4, J=200)$ with the normality assumption of $C_{1j}$ violated.}
\begin{adjustbox}{max width=\textwidth}
\renewcommand{\arraystretch}{1.5}
    \begin{tabular}{cccccccccc}
\toprule
  & \multicolumn{3}{c}{CDML} & \multicolumn{3}{c}{GSExact} & \multicolumn{3}{c}{Blimp} \\
\cmidrule(lr){2-4}\cmidrule(lr){5-7}\cmidrule(lr){8-10}
 Simulated & \%Bias (ASE) & ESE & Coverage & \%Bias (ASE) & ESE & Coverage  & \%Bias (ASE) & ESE & Coverage \\ \midrule
 
 $\tau$=4 & 0.0 (0.84) & 0.87 & 0.94 & -3.2 (1.08) & 1.13  & 0.93 & 4.4 (1.15) & 1.17 & 0.93 \\

 $\sigma^2$=16 & -0.2 (0.92) & 0.92 & 0.95 & 0.5 (1.08) & 1.08 & 0.95 & 0.2 (1.08) & 1.07  & 0.96  \\

 $\beta_{0}$=1 & 1.7 (0.99) & 1.03 & 0.94 & 8.1 (1.17) & 1.22 & 0.93 & 10.7 (1.19) & 1.22 & 0.94 \\

 $\beta_{1}$=1 & -0.8 (0.45) & 0.48 & 0.93 & -5.6 (0.54) & 0.58 & 0.92 & -6.5 (0.55) & 0.58 & 0.93 \\

 $\beta_{2}$=1 & 0.0 (0.42) & 0.44 & 0.93 & -3.2 (0.51) & 0.53 & 0.94 & -3.9 (0.52) & 0.53 & 0.94 \\

 $\beta_{3}$=1 & -0.3 (0.21)  & 0.21 & 0.96 & -0.2 (0.25) & 0.25  & 0.96 & 0.3 (0.26) & 0.25 & 0.96 \\

 $\beta_{4}$=1 & 0.3 (0.18)  & 0.20 & 0.94 & 2.5 (0.22) & 0.24  & 0.92 & 2.5 (0.23) & 0.24 & 0.93\\
\bottomrule
\end{tabular}%
\end{adjustbox}
\label{robust200}%
\end{table}%

Table \ref{robust200} summarizes the estimated HLM for the large-sample case. Because the CDML estimation given complete data does not depend on the distributional assumption of $C_{1j}$, the CDML biases and ASEs near ESEs are small with good coverages. Both GSExact and Blimp result in reasonably small biases up to 4.4\% in magnitude, except the 8.1\% and 10.7\% biases in the intercept and the -5.6\% and -6.5\% biases in the effect $\beta_1$ of $C_{1j}$, respectively. They also produce reasonably small and similar ASEs and ESEs, with similar coverages near the nominal level. Overall, both GSExact and Blimp produce estimates that are quite robust against the violated normality assumption in the large-sample scenario. 

\begin{table}[htbp]
\centering
  \caption{\textcolor{black}{Estimated \% biases, ASEs, ESEs and coverages from 5,000 simulated data sets of the large sample size $(n_j=4, J=36)$ with the normality assumption of $C_{1j}$ violated.}}
\begin{adjustbox}{max width=\textwidth}
\renewcommand{\arraystretch}{1.5}
    \begin{tabular}{cccccccccc}
\toprule
  & \multicolumn{3}{c}{CDML} & \multicolumn{3}{c}{GSExact} & \multicolumn{3}{c}{Blimp} \\
\cmidrule(lr){2-4}\cmidrule(lr){5-7}\cmidrule(lr){8-10}
 Simulated & \%Bias (ASE) & ESE & Coverage & \%Bias (ASE) & ESE & Coverage  & \%Bias (ASE) & ESE & Coverage \\ \midrule
 
 $\tau$=4 & 0.0 (2.10) & 2.06 & 0.96& 1.2 (2.65) & 2.10  & 0.97 & 35.7 (3.88) & 2.89 & 0.96 \\

 $\sigma^2$=16 & -0.2 (2.17) & 2.15 & 0.95 & 1.5 (2.56) & 2.47 & 0.95 & 2.0 (2.66) & 2.48  & 0.95  \\

 $\beta_{0}$=1 & -5.3 (2.67) & 2.78 & 0.94 & 10.6 (3.62) & 3.77 & 0.94 & 23.9 (3.98) & 3.70 & 0.96 \\

 $\beta_{1}$=1 & 3.8 (1.27) & 1.33 & 0.94 & -9.0 (1.75) & 1.84 & 0.93 & -11.9 (1.93) & 1.81 & 0.96 \\

 $\beta_{2}$=1 & 2.3 (1.17) & 1.23 & 0.94 & -8.1 (1.62) & 1.69 & 0.93 & -10.9 (1.78) & 1.66 & 0.95 \\

 $\beta_{3}$=1 & -0.4 (0.54)  & 0.55 & 0.95 & 1.0 (0.70) & 0.71  & 0.94 & 3.6 (0.77) & 0.71 & 0.97 \\

 $\beta_{4}$=1 & -1.4 (0.54)  & 0.57 & 0.94 & 5.0 (0.76) & 0.81  & 0.93 & 4.3 (0.83) & 0.80 & 0.95\\
\bottomrule
\end{tabular}%
\end{adjustbox}
\label{robust36}%
\end{table}%

Table \ref{robust36} summarizes the simulation results for the small sample size. Almost all biases, ASEs and ESEs are larger than those of the large-sample results in Table \ref{robust200}. The CDML biases are still reasonably small up to -5.3\% in magnitude with small ASEs near ESEs and good coverages. GSExact estimates result in biases of 10.6\%, -9.0\% and -8.1\% for the intercept $\beta_0$, the effect $\beta_1$ of $C_{1j}$ and the effect $\beta_2$ of $C_{2j}$, respectively. Other estimates are reasonably accurate with biases up to 5\%. ASEs are modestly distant from, but reasonably close to ESEs, with coverages near the nominal level. Blimp estimates produce higher biases than their GSExact counterparts overall. The biases for the intercept, the effect of $C_{1j}$ and the effects of $C_{2j}$ are 23.9\%, -11.9\% and -10.9\%, respectively, which are noticeably higher than the GSExact counterparts. The $\tau$ estimator is biased by 37\% and is the least inaccurate. Other biases are below 4.3\%. ASEs are generally higher than ESEs and their GSExact counterparts. The coverages are near the nominal level. Therefore, GSExact estimators appear more robust than their Blimp counterparts overall under our scenario involving the violated normality assumption of a covariate and a small sample size.   

\textcolor{black}{\subsection{Violation of the MAR Assumption}}
To assess the robustness of our estimators against violations of the MAR assumption, we designed a simulation that incorporates both MAR and MNAR (Missing Not At Random) mechanisms \citep{rubin1976inference}. After simulating sample data as in Section 4.1, we simulate the missing values of $C_{1j}$ and $C_{2j}$ using a $Bernoulli(p_j)$ under a MNAR mechanism that depends on the values of $C_{1j}$ (before simulating missing values) where
\begin{align*}
logit(p_j) = d_0+d_1C_{1j}. 
\end{align*}
\noindent To simulate missing values, we set $d_0=-5$ and $ d_1=1.3$ for $C_{1j}$, and $d_0=-10.5$ and $ d_1=3$ for $C_{2j}$. These parameters are chosen to produce approximately 20\% missing values for each covariate. This approach allows us to simulate data under a MNAR mechanism, as the missing pattern of $C_{1j}$ and $C_{2j}$ depends on the missing values of $C_{1j}$. We maintain the MAR mechanism (\ref{missing_mechanism}) to simulate the missing values of $Y_{ij}$, as described in Section 4.1. We evaluate our estimators under this combination of the MAR and MNAR mechanisms by simulating small sample sizes.

\begin{table}[htbp]
\centering
  \caption{\textcolor{black}{Estimated \% biases, ASEs, ESEs and coverages from 5,000 small-sample simulations $(n_j=4, J=36)$ with the MNAR mechanism for $C_{1j}$ and $C_{2j}$ and the MAR mechanism for $Y_{ij}$}}
\begin{adjustbox}{max width=\textwidth}
\renewcommand{\arraystretch}{1.5}
    \begin{tabular}{cccccccccc}
\toprule
  & \multicolumn{3}{c}{CDML} & \multicolumn{3}{c}{GSExact} & \multicolumn{3}{c}{Blimp} \\
\cmidrule(lr){2-4}\cmidrule(lr){5-7}\cmidrule(lr){8-10}
 Simulated & \%Bias (ASE) & ESE & Coverage & \%Bias (ASE) & ESE & Coverage  & \%Bias (ASE) & ESE & Coverage \\ \midrule
 
 $\tau$=4 & 1.0 (2.98) & 2.12 & 0.95 & 4.7 (2.73) & 2.21  & 0.97 & 39.4 (4.01) & 3.0 & 0.95 \\

 $\sigma^2$=16 & -0.2 (2.96) & 2.17 & 0.96 & 1.6 (2.63) & 2.49 & 0.95 & 2.1 (2.71) & 2.50 & 0.95  \\

 $\beta_{0}$=1 & 0.1 (1.51) & 1.58 & 0.93 & 5.1 (1.91) & 1.90 & 0.95 & -5.4 (2.12) & 1.89 & 0.97 \\

 $\beta_{1}$=1 & -0.9 (0.68) & 0.72 & 0.94 & -4.3 (1.00) & 1.01 & 0.94 & -1.4 (1.11) & 1.00 & 0.96 \\

 $\beta_{2}$=1 & 1.1 (0.87) & 0.91 & 0.94 & 2.4 (1.13) & 1.16 & 0.94 & 5.5 (1.25) & 1.14 & 0.96 \\

 $\beta_{3}$=1 & 0.7 (0.94)  & 0.96 & 0.94 & 2.6 (1.20) & 1.21  & 0.94 & 10.1 (1.34) & 1.20  & 0.96 \\

 $\beta_{4}$=1 & -0.2 (0.30)  & 0.31 & 0.95 & 4.0 (0.47) & 0.47  & 0.94 & -0.5 (0.52) & 0.46  & 0.97`\\
\bottomrule
\end{tabular}%
\end{adjustbox}
\label{mnar5k}%
\end{table}%

Table \ref{mnar5k} summarizes the resulting estimates. The CDML estimates remain unaffected by the violation, resulting in accurate and precise estimates. In contrast, the GSExact biases for the fixed effects range from 2.4\% to 5.1\% in magnitude, which are comparatively larger than the CDML counterparts. Additionally, bias for the random effect $\tau$ is 4.7\%, approximately four times larger than that of the CDML counterpart. The ASEs for GSExact are a slightly higher or lower than those of CDML, and are reasonably close to the ESEs. Coverages are near the nominal 0.95.

The Blimp biases for the variance $\tau$ and the effect $\beta_3$ of $X_j$ are 39.4\% and 10.1\%, larger than their GSExact counterparts (4.7\% and 2.6\%). Other biases are modest, up to 5.5\%, and moderately larger or smaller than the GSExact counterparts. ASEs are reasonably close to ESEs, but noticeably larger than the GSExact ASEs, with coverages near the nominal level.

Overall, GSExact demonstrated comparative robustness, with most biases up to 5.1\% under the simulated MNAR scenario. The Blimp estimators of $\tau$, the main effect $\beta_2$ of $C_{2j}$ and the main effect $\beta_3$ of $X_j$ were less robust than the GSExact estimators, while the remaining Blimp estimators exhibited comparable robustness to their GSExact counterparts when the MAR assumption was violated in small sample sizes. These findings, however, are based on a single MNAR mechanism we simulated. Investigating robust estimation under different violations of the MAR assumption is an important avenue for future research, which is beyond the scope of this paper.

\textcolor{black}{\subsection{Controlling for More Covariates or Nonlinear Terms}}
We now evaluate the performance of our estimators while controlling for additional covariates. In particular, we simulate data as in Section 4.1 with the HLM (\ref{HLM1}): $Y_{ij} \sim N(1 + C_{1j} + C_{2j} + X_j + C_{1j}C_{2j} +C_{1j}X_j + C_{2j}X_j, \tau + \sigma^2)$, where $\tau=4$ and $\sigma^2=16$. This model includes two additional interaction terms, $C_{1j}X_j$ and $C_{2j}X_j$, which are partially observed. We then generate missing values in $Y_{ij}$, $C_{1j}$, and $C_{2j}$ based on the MAR mechanism (\ref{missing_mechanism}). Again, we simulate the partially observed data with small sample sizes 5,000 times.

\begin{table}[htbp]
\centering
  \caption{\textcolor{black}{Estimated \% biases, ASEs, ESEs, and coverages from 5,000 simulated small-sample data sets $(n_j=4, J=36)$ with $C_{1j}$, $C_{2j}$, and $Y_{ij}$ MAR}}
\begin{adjustbox}{max width=\textwidth}
\renewcommand{\arraystretch}{1.5}
    \begin{tabular}{cccccccccc}
\toprule
  & \multicolumn{3}{c}{CDML} & \multicolumn{3}{c}{GSExact} & \multicolumn{3}{c}{Blimp} \\
\cmidrule(lr){2-4}\cmidrule(lr){5-7}\cmidrule(lr){8-10}
 Simulated & \%Bias (ASE) & ESE & Coverage & \%Bias (ASE) & ESE & Coverage  & \%Bias (ASE) & ESE & Coverage \\ \midrule
 
 $\tau$=4 & -0.3 (2.17) & 2.17 & 0.96 & -0.2 (2.73) & 2.21  & 0.97 & 37.2 (4.01) & 3.0 & 0.95 \\

 $\sigma^2$=16 & 0.1 (2.18) & 2.17 & 0.95 & 1.6 (2.63) & 2.49 & 0.95 & 2.3 (2.71) & 2.50 & 0.95  \\

 $\beta_{0}$=1 & -1.5 (2.07) & 2.21 & 0.94 & -1.9 (1.91) & 1.90 & 0.95 & 7.5 (2.12) & 1.89 & 0.97 \\

 $\beta_{1}$=1 & 1.6 (1.06) & 1.10 & 0.94 & 2.2 (1.00) & 1.01 & 0.94 & -8.4 (1.11) & 1.00 & 0.96 \\

 $\beta_{2}$=1 & -0.3 (1.23) & 1.30 & 0.94 & -3.6 (1.13) & 1.16 & 0.94 & -5.8 (1.25) & 1.14 & 0.96 \\

 $\beta_{3}$=1 & -0.3 (1.71)  & 1.79 & 0.94 & -0.3 (1.20) & 1.21  & 0.94 & 5.3 (1.34) & 1.20  & 0.96 \\

 $\beta_{4}$=1 & 0.4 (0.48)  & 0.52 & 0.94 & 2.4 (0.47) & 0.47  & 0.94 & 1.1 (0.52) & 0.46  & 0.97\\

 $\beta_{5}$=1 & -0.6 (0.57)  & 0.62 & 0.94 & -2.3 (0.47) & 0.47  & 0.94 & 1.3 (0.52) & 0.46  & 0.97\\

 $\beta_{6}$=1 & 0.2 (0.42)  & 0.45 & 0.94 & 1.3 (0.47) & 0.47  & 0.94 & 0.3 (0.52) & 0.46  & 0.97\\
\bottomrule
\end{tabular}%
\end{adjustbox}
\label{moreinter}%
\end{table}%

Table \ref{moreinter} summarizes the estimates. The CDML estimates are accurate and precise as before, with modestly increased ASEs from those in Table \ref{j=36} to reflect more parameters estimated. The GSExact estimators are biased by up to -3.6\% in magnitude, maintaining the accuracy of those in Table \ref{j=36} despite the addition of two partially observed nonlinear terms. ASEs and ESEs are slightly increased comparatively, reflecting increased uncertainty due to additional partially observed interaction terms in the HLM, and coverages are good near the nominal level.

The Blimp bias for $\tau$ is 37.2\%, higher than the 25.9\% bias in Table \ref{j=36}, while other biases change less noticeably. The Blimp biases are noticeably higher than those of the GSExact estimators, except for the biases for the interaction effects, which are lower than their GSExact counterparts. Blimp ASEs are also modestly inflated compared to the GSExact ASEs. 

These differences in accuracy and precision of GSExact and Blimp estimators highlight the importance of GSExact implementing exact posteriors and, thus, ensuring compatibility. In contrast, Blimp implements a Gibbs sampler using a Metropolis algorithm, which has not been shown to ensure compatibility. Confer Shin and Raudenbush\citep{shin2024b} for a practical implication of compatibility.
\\
\\
\section{Analysis of Racially Discordant Patient-Physician Interactions}
We analyze data from racially discordant medical encounters between patients and physicians by GSExact. Investigators video-recorded physicians' behavioral and facial expressions during medical interactions with patients in office visits, and coded physicians' communication behaviors from the recordings. Each physician completed a baseline survey on demographics and other characteristics including implicit prejudice measured by the Implicit Association Test \citep{greenwald1998measuring,nosek2005understanding} and explicit prejudice measured by a symbolic racism test 2000 \citep{henry2002symbolic}. Each encounter lasted about 20 minutes, and the video-taped physician's facial expression of each encounter was rated by a machine as a valence score at each of four consecutive time points. Due to the COVID-19 restrictions, the investigators were able to recruit only 37 patients and 6 physicians, much less than planned.  

Of interest are the main and interaction effects of two physician characteristics, implicit prejudice($IPre$) and explicit prejudice($EPre$), on their communication behavior or a positive valence score ($Valence$) that measures the intensity of positive facial expression during the medical encounter. These variables are partially observed. We also analyze a fully known physician's Communication Training($CT$) covariate as the time elapsed since the last training on communication skills. Because a positive valence score is expected to be higher at the beginning and end (e.g., when greeting) than mid time points of the encounter (e.g., when talking about health issues), we also control for dummy variables indicating the second to fourth measurement time points of the outcome Q2, Q3 and Q4 with the first time point as a reference. To improve the interpretability of the intercept and reduce potential collinearity between the covariates and the interaction term, we center the covariates at their sample means and write the model
\begin{align}
\label{realHLM}
Valence_{ij}=\beta_0+\beta_1IPre_j+\beta_2EPre_j+\beta_3CT_j+\beta_4IPre_jEPre_j+u_j+\beta_5Q_{2ij}+\beta_6Q_{3ij}+\beta_7Q_{4ij}+e_{ij},
\end{align}
\noindent where occasion $i$ is nested within the $j$th patient-physician encounter for $i=1,\cdots,4$ and $j=1,\cdots,37$. Here  $\beta_0$ is the expected positive valence score of encounters for physicians having average implicit prejudice ($IPre$), average explicit prejudice ($EPre$) and average level of communication training ($CT$) at occasion 1 ($Q_2=Q_3=Q_4=0$). The fixed effects include the main effects $\beta_1$, $\beta_2$ and $\beta_3$ of $IPre$, $EPre$ and $CT$, respectively; the interaction effect $\beta_4$ of $IPre$ and $EPre$; and the mean differences $\beta_5$, $\beta_6$ and $\beta_7$ from the mean positive valence of occasion one at occasions 2-4, respectively, ceteris paribus. An encounter-specific random effect $u_j\sim \mathcal{N}(0,\tau)$ and an occasion-specific random effect $e_{ij} \sim N(0,\sigma^2)$ are independent. The valence score is missing 20\% of the values, and one physician failed to report both $Ipre$ and $Epre$ such that 16\% of their values are missing, respectively.

\begin{table}[htbp]
    \centering
    \caption{Estimated HLM (\ref{realHLM}) for analysis of racially discordant patient-physician encounters.}
    \renewcommand{\arraystretch}{1.5}
    \begin{tabular}{cccc}
        \toprule
        Parameters & Covariates & \multicolumn{1}{c}{Estimates (se\textsuperscript{+})} & \multicolumn{1}{c}{CI\textsuperscript{++} (2.5th \%ile, 97.5th \%ile)} \\
        \midrule
        $\beta_{0}$ & Intercept & 84.46(2.42)\textsuperscript{*} & (79.71, 107.21) \\
        $\beta_{1}$ & IPrej & -0.35(5.62) & (-11.29, 10.72) \\
        $\beta_{2}$ & EPrej & -21.83(11.22)\textsuperscript{*} & (-44.61, -1.49) \\
        $\beta_{3}$ & CT & -0.65(1.81) & (-3.99, 3.24) \\
        $\beta_{4}$ & IPrej$\times$EPrej & -78.24(52.36) & (-188.84, 14.82) \\
        $\beta_{5}$ & Q2 & -9.58(2.65)\textsuperscript{*} & (-14.99, -4.62) \\
        $\beta_{6}$ & Q3 & -4.25(2.59) & (-9.30, 0.68) \\
        $\beta_{7}$ & Q4 & -2.14(2.47) & (-7.10, 2.66) \\
        $\tau$ & - & 4.57(5.07) & (0.57, 19.59) \\
        $\sigma^2$ & - & 80.89(12.48) & (59.33, 107.21) \\
        \bottomrule
    \end{tabular}%
    \label{realdata}%
    \caption*{\footnotesize{+: standard error; ++: a 95\% credible interval; *: Significantly different from zero at a level 0.05}}
\end{table}%

Table \ref{realdata} shows the estimated HLM by GSexact. Each estimate and its associated standard error in parentheses are listed in column three followed by a 95\% Bayesian credible interval in the last column. At a significance level 0.05, explicit prejudice is negatively associated with positive valence score ($\beta_2$=-21.83, standard error(se)=11.22), and the mean outcome at occasion two is 9.58 (se=2.65) units less than that at occasion one, controlling for other covariates in the model. Neither large antagonistic interaction effect -78.24 (52.36) of implicit prejudice and explicit prejudice nor the effect -0.35 (5.62) of implicit prejudice is statistically significant, ceteris paribus. The intra-cluster correlation coefficient 4.57/(4.57+80.89)=0.05 implies that only 5\% of the remaining total variance in the positive valence score has to be explained at the encounter level. Therefore, high explicit prejudice, measured on the symbolic racism scale, is associated with low positive valence score during patient-physician medical encounters on average. Additionally, the mean positive valence score at occasion 2 drops significantly from the mean initial score at occasion 1, controlling for other covariates in the model. Finally, we used 2,500 burn-in iterations and 2,500 post burn-in iterations. The PSRF statistic for each parameter was less than 1.1, satisfying the convergence criterion. \textcolor{black}{The R package GSExact, available at https://github.com/shind10/GSExact, provides all simulation codes from Section 4, as well as the estimation in Table \ref{realdata}}. 

\section{Discussion}
We estimated a HLM with the nonlinear effects of cluster-level covariates from partially observed data assumed missing at random. Our Bayesian estimation by a Gibbs sampler consists of exact posterior distributions that ensure compatibility with the HLM. Our simulation results show that the sampler produces reasonably accurate and precise estimates given sample sizes as small as $n=4$ units nested within $J=36$ clusters and given partially observed outcome and cluster-level continuous covariates having interaction effects. Our estimators were as accurate and precise as those of competing methods given large sample sizes $(n=4, J=200)$, and more accurate and precise than completing ones given small sample sizes $(n=4, J=36)$ in our simulation study.

We checked convergence to the posterior distribution by the potential scale reduction factor (PSRF) and Geweke's criteria, and found that the PSRF criterion was preferable to Geweke's criterion in our simulation scenarios. This preference was more apparent given additional partially observed covariates and smaller sample sizes.

\textcolor{black}{In our simulation, we also assessed robustness of our estimators under three cases: i) a misspecified distribution of a partially observed covariate; ii) violation of the MAR assumption; and iii) multiple interaction terms of partially observed covariates.} The simulation results reveal that our estimators are reasonably robust against these violations given large and small sample sizes in our simulation scenarios. An importance avenue for future research is to conduct extensive simulations under different misspecified models and missing data mechanisms.

Our simulation study is limited in the sense that the cluster size was fixed at $n=4$ units, as analyzed in our real data analysis, while the number of $J$ clusters were varied. Valuable future research is to assess the impact of varied cluster sizes on our estimators. 

In near future, we aim to extend our sampler to efficiently handle partially observed discrete covariates or a mixture of partially observed discrete and continuous covariates having interaction effects at the cluster level. We also plan to extend our method to models where partially observed lower-level covariates have nonlinear effects. See Shin and Raudenbush \citep{shin2024maximum} for ML estimation of a HLM for covariates MAR with nonlinear effects.

%%%%%%%%%%%%%%%%%%%%%%%%%%%%%%%%%%%%%%%%%%%%%%%%%%%%
%\clearpage

\clearpage
%\section*{Acknowledgements}
%Acknowledgements
\bibliography{reference}

\textcolor{black}{\section*{Appendix A : Additional Trace plots}}

\begin{figure}[ht]
\centering
    \begin{minipage}[b]{0.48\textwidth}
        \centering
        \includegraphics[width=\textwidth]{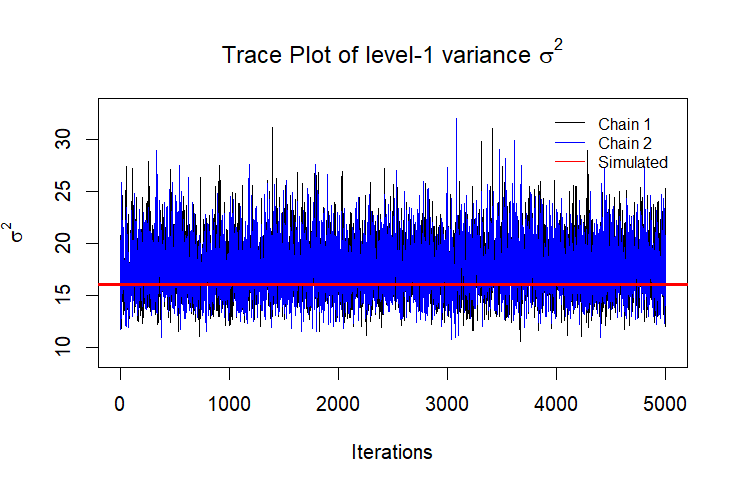}
    \end{minipage}
\hfill
    \begin{minipage}[b]{0.48\textwidth}
        \centering
        \includegraphics[width=\textwidth]{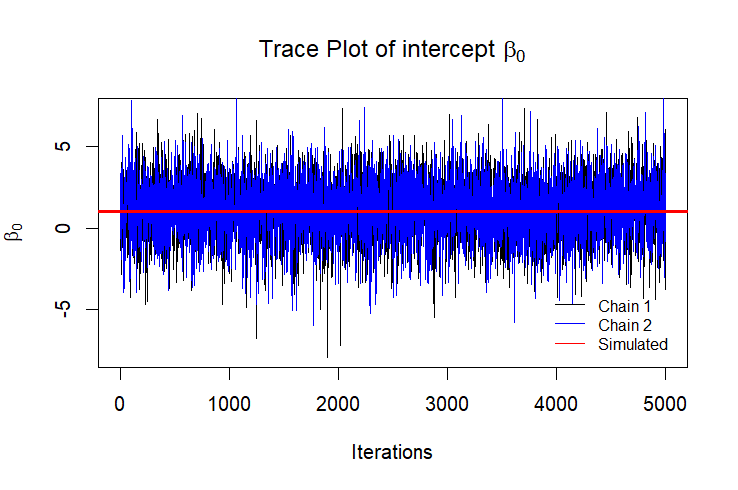}
    \end{minipage}
\\
    \begin{minipage}[b]{0.48\textwidth}
        \centering
        \includegraphics[width=\textwidth]{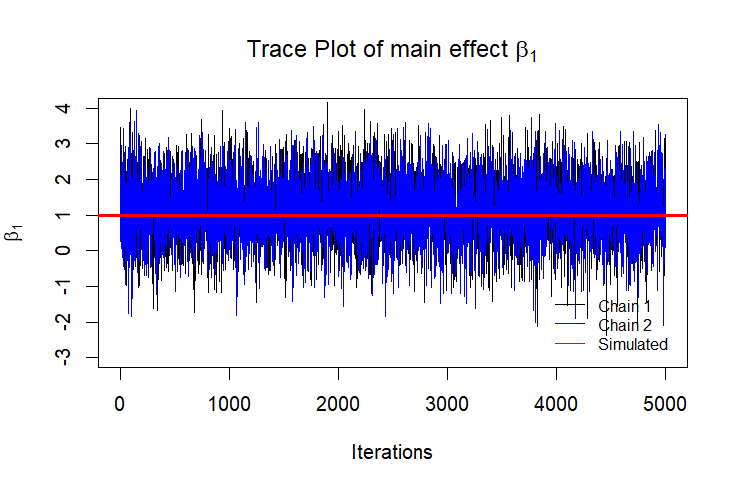}
    \end{minipage}
\hfill
    \begin{minipage}[b]{0.48\textwidth}
        \centering
        \includegraphics[width=\textwidth]{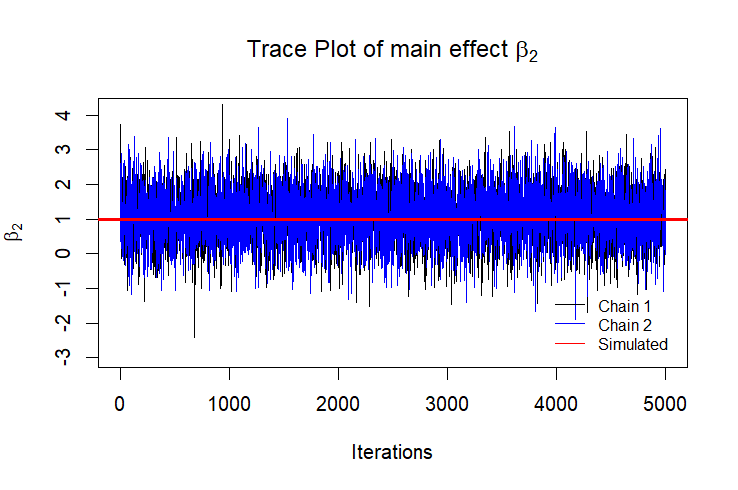}
    \end{minipage}
\\
    \begin{minipage}[b]{0.48\textwidth}
        \centering
        \includegraphics[width=\textwidth]{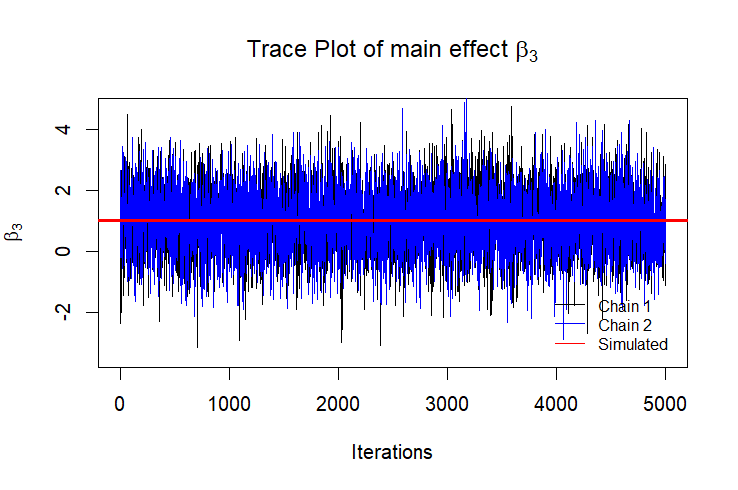}
    \end{minipage}
\caption{Trace plots for other parameters}
\label{fig:traceplots_other}
\end{figure}

Figure \ref{fig:traceplots_other} presents additional trace plots for parameters, not discussed in the main text. As described in the simulation study section, these trace plots indicate converged chains with rapid random fluctuations around a stable confidence band, no discernible patterns, and extensive overlap between chains.

\textcolor{black}{\section*{Appendix B : GSExact Versus Imputation Using Predictive Mean Matching}}

Multiple Imputation by Chained Equations (MICE) based on the Fully Conditional Specification (FCS) algorithm is statistically incompatible with an analytic HLM that include nonlinear effects. This incompatibility arises because MICE uses an incompatible imputation model, leading to biased estimation (Seaman et al., 2012; Liu et al., 2014; Kim et al., 2015; Bartlett et al., 2015; Enders et al., 2020).

Compared to parametric approaches for specifying imputation models, imputation using predictive mean matching (PMM) may be robust against misspecified imputation models because PMM preserves the distribution of the original data by imputing a missing value as an observed one that is closest in predictive means. PMM, however, becomes problematic in small samples, as analyzed in this paper, where observed donors are scarce. This can lead to a lack of variability in the imputed values and potentially biased estimates, especially for higher-level variables in multilevel data.

To illustrate this, we implemented PMM in the MICE package to impute the missing values for the same 5,000 simulated small data sets that we used to compare performance between CDML, GSExact, and BLIMP in Table \ref{j=36}, where we replaced the CDML estimates with the PMM results.

\begin{table}[htbp]
\centering
  \caption{Estimated \% biases, ASEs, ESEs and coverages from 5,000 simulated data sets from the small sample simulation $(n_j=4, J=36)$ with missing values of $C_{1j}$, $C_{2j}$ and $Y_{ij}$ are based on MAR.}
\begin{adjustbox}{max width=\textwidth}
\renewcommand{\arraystretch}{1.5}
    \begin{tabular}{cccccccccc}
\toprule
  & \multicolumn{3}{c}{MICE(PMM)} & \multicolumn{3}{c}{GSExact} & \multicolumn{3}{c}{Blimp} \\
\cmidrule(lr){2-4}\cmidrule(lr){5-7}\cmidrule(lr){8-10}
 Simulated & \%Bias (ASE) & ESE & Coverage & \%Bias (ASE) & ESE & Coverage  & \%Bias (ASE) & ESE & Coverage \\ \midrule
 
 $\tau$=4 & 12.0 (2.29) & 2.66 & 0.92 & -1.9 (2.54) & 2.00  & 0.95 & 25.9 (3.57) & 2.60 & 0.97 \\

 $\sigma^2$=16 & 5.4 (2.31) & 3.16 & 0.85 & 1.9 (2.57) & 2.49 & 0.94 & 2.5 (2.66) & 2.51 & 0.95  \\

 $\beta_{0}$=1 & -130.3 (1.58) & 2.07 & 0.81 & 6.3 (1.89) & 1.84 & 0.97 & 4.1 (2.16) & 1.90 & 0.97 \\

 $\beta_{1}$=1 & 52.3 (0.67) & 0.96 & 0.76 & -0.9 (0.90) & 0.88 & 0.97 & -4.7 (1.02) & 0.90 & 0.96 \\

 $\beta_{2}$=1 & 53.9 (0.87) & 1.16 & 0.81 & -0.9 (1.13) & 1.13 & 0.96 & -5.9 (1.26) & 1.11 & 0.97 \\

 $\beta_{3}$=1 & 28.8 (0.91)  & 1.42 & 0.79 & -3.8 (1.19) & 1.20  & 0.97 & 8.9 (1.31) & 1.19  & 0.96 \\

 $\beta_{4}$=1 & -35.8 (0.30)  & 0.35 & 0.71 & 1.3 (0.40) & 0.39  & 0.96 & 0.3 (0.44) & 0.39  & 0.97\\
\bottomrule
\end{tabular}%
\end{adjustbox}
\label{AppendixTable}%
\end{table}%

Table \ref{AppendixTable} summarizes the results. PMM biases are larger than those of GXExact and Blimp overall. Given the small number of 36 clusters, the scarcity of observed donors limits the ability to  impute cluster-level missing values, leading to underestimated standard errors.

\end{document}

%% file: preamble.tex
\usepackage{comment} 
\usepackage{longtable}
\usepackage{adjustbox}
\usepackage{graphicx}

\usepackage[caption = false]{subfig}
\usepackage{enumitem}
\usepackage{amsmath}
\usepackage{textcomp}
\usepackage{amsfonts}
\usepackage{pdfpages}
\usepackage{verbatim}
\usepackage{natbib}
\usepackage{enumitem}
\usepackage{enumerate}
\usepackage{setspace}

\usepackage{lineno,dcolumn, booktabs,multirow, esvect,xcolor, colortbl,changepage}
\usepackage{multirow}
\usepackage{supertabular}
\usepackage{booktabs}
\usepackage{epstopdf}
\usepackage{srcltx,epsfig,bm,color,rotating}
\usepackage{lscape}
\usepackage{listings}
\usepackage{booktabs,caption}
\usepackage[flushleft]{threeparttable}

\allowdisplaybreaks
\usepackage{float}

\usepackage{url}
\usepackage{xcite}
\usepackage{setspace}
\usepackage{subfig}
\usepackage{graphicx}

\usepackage{hyperref}

\graphicspath{ {image/} }

%% file: ShinDH.bbl
\begin{thebibliography}{10}
\providecommand \doibase [0]{http://dx.doi.org/}%

\bibitem{raudenbush2002hierarchical}
Raudenbush SW, Bryk AS. {\it Hierarchical linear models: Applications and data analysis methods}. 1.
\newblock sage .
\newblock 2002.

\bibitem{goldstein2011multilevel}
Goldstein H. {\it Multilevel statistical models}.
\newblock John Wiley \& Sons .
\newblock 2011.

\bibitem{little2002bayes}
Little RJ, Rubin DB. Bayes and multiple imputation. {\it Statistical analysis with missing data} 2002\string: 200--220.

\bibitem{rubin1987}
Rubin DB. {\it Multiple Imputation for Nonresponse in Surveys}.
\newblock John Wiley \& Sons, Inc. .
\newblock 1987.

\bibitem{rubin1976inference}
Rubin DB. Inference and missing data. {\it Biometrika} 1976\string; 63(3)\string: 581--592.

\bibitem{schafer2002computational}
Schafer JL, Yucel RM. Computational strategies for multivariate linear mixed-effects models with missing values. {\it Journal of computational and Graphical Statistics} 2002\string; 11(2)\string: 437--457.

\bibitem{shin2007just}
Shin Y, Raudenbush SW. Just-identified versus overidentified two-level hierarchical linear models with missing data. {\it Biometrics} 2007\string; 63(4)\string: 1262--1268.

\bibitem{raghunathan2001multivariate}
Raghunathan TE, Lepkowski JM, Van~Hoewyk J, Solenberger P, others . A multivariate technique for multiply imputing missing values using a sequence of regression models. {\it Survey methodology} 2001\string; 27(1)\string: 85--96.

\bibitem{van2007multiple}
Van~Buuren S. Multiple imputation of discrete and continuous data by fully conditional specification. {\it Statistical methods in medical research} 2007\string; 16(3)\string: 219--242.

\bibitem{van2011mice}
Van~Buuren S, Groothuis-Oudshoorn K. mice: Multivariate imputation by chained equations in R. {\it Journal of statistical software} 2011\string; 45\string: 1--67.

\bibitem{van2018flexible}
Van~Buuren S. {\it Flexible imputation of missing data}.
\newblock CRC press .
\newblock 2018.

\bibitem{arnold1989compatible}
Arnold BC, Press SJ. Compatible conditional distributions. {\it Journal of the American Statistical Association} 1989\string; 84(405)\string: 152--156.

\bibitem{arnold2001conditionally}
Arnold BC, Castillo E, Sarabia JM. Conditionally specified distributions: an introduction (with comments and a rejoinder by the authors). {\it Statistical Science} 2001\string; 16(3)\string: 249--274.

\bibitem{liu2014stationary}
Liu J, Gelman A, Hill J, Su YS, Kropko J. On the stationary distribution of iterative imputations. {\it Biometrika} 2014\string; 101(1)\string: 155--173.

\bibitem{bartlett2015multiple}
Bartlett JW, Seaman SR, White IR, Carpenter JR, Initiative* ADN. Multiple imputation of covariates by fully conditional specification: accommodating the substantive model. {\it Statistical methods in medical research} 2015\string; 24(4)\string: 462--487.

\bibitem{enders2016multilevel}
Enders CK, Mistler SA, Keller BT. Multilevel multiple imputation: A review and evaluation of joint modeling and chained equations imputation.. {\it Psychological methods} 2016\string; 21(2)\string: 222.

\bibitem{kim2015evaluating}
Kim S, Sugar CA, Belin TR. Evaluating model-based imputation methods for missing covariates in regression models with interactions. {\it Statistics in medicine} 2015\string; 34(11)\string: 1876--1888.

\bibitem{shin2024maximum}
Shin Y, Raudenbush SW. Maximum Likelihood Estimation of Hierarchical Linear Models from Incomplete Data: Random Coefficients, Statistical Interactions, and Measurement Error. {\it Journal of Computational and Graphical Statistics} 2024\string; 33(1)\string: 112--125.

\bibitem{carpenterandkenward}
Carpenter J, Kenward M. {\it Multiple Imputation and Its Application}.
\newblock John Wiley \& Sons Inc .
\newblock 2013.

\bibitem{erler2016dealing}
Erler NS, Rizopoulos D, Rosmalen Jv, Jaddoe VW, Franco OH, Lesaffre EM. Dealing with missing covariates in epidemiologic studies: a comparison between multiple imputation and a full Bayesian approach. {\it Statistics in medicine} 2016\string; 35(17)\string: 2955--2974.

\bibitem{enders2020model}
Enders CK, Du H, Keller BT. A model-based imputation procedure for multilevel regression models with random coefficients, interaction effects, and nonlinear terms.. {\it Psychological methods} 2020\string; 25(1)\string: 88.

\bibitem{kim2018multiple}
Kim S, Belin TR, Sugar CA. Multiple imputation with non-additively related variables: Joint-modeling and approximations. {\it Statistical methods in medical research} 2018\string; 27(6)\string: 1683--1694.

\bibitem{goldstein2014fitting}
Goldstein H, Carpenter JR, Browne WJ. Fitting multilevel multivariate models with missing data in responses and covariates that may include interactions and non-linear terms. {\it Journal of the Royal Statistical Society Series A: Statistics in Society} 2014\string; 177(2)\string: 553--564.

\bibitem{metropolis1953equation}
Metropolis N, Rosenbluth AW, Rosenbluth MN, Teller AH, Teller E. Equation of state calculations by fast computing machines. {\it The journal of chemical physics} 1953\string; 21(6)\string: 1087--1092.

\bibitem{hastings1970monte}
HASTINGS W. Monte Carlo sampling methods using Markov chains and their applications. {\it Biometrika} 1970\string; 57(1)\string: 97--97.

\bibitem{shin2024b}
Shin Y, Raudenbush SW. Multiple Imputation to Estimate Hierarchical Models from Data MAR: Latent Covariates, Random Coefficients and Statistical Interactions. {\it Submitted} 2024.

\bibitem{hoff2009first}
Hoff PD. {\it A first course in Bayesian statistical methods}. 580.
\newblock Springer .
\newblock 2009.

\bibitem{douglas2015fitting}
Douglas~Bates M, Bolker B, Walker S. Fitting linear mixed-effects models using lme4. {\it Journal of Statistical Software} 2015\string; 67(1)\string: 1--48.

\bibitem{blimp}
Keller B, Enders C. Blimp user's guide(Version 3) Retrieved from www.appliedmissingdata.com/blimp.  2022.

\bibitem{geweke1991evaluating}
Geweke J. Evaluating the accuracy of sampling-based approaches to the calculation of posterior moments. tech. rep., Federal Reserve Bank of Minneapolis; Minneapolis, MN:   1991.

\bibitem{brooks1998general}
Brooks SP, Gelman A. General methods for monitoring convergence of iterative simulations. {\it Journal of computational and graphical statistics} 1998\string; 7(4)\string: 434--455.

\bibitem{du2022performances}
Du H, Ke Z, Jiang G, Huang S. The Performances of Gelman-Rubin and Geweke's Convergence Diagnostics of Monte Carlo Markov Chains in Bayesian Analysis. {\it Journal of Behavioral Data Science} 2022\string; 2(2)\string: 47--72.

\bibitem{cowles1996markov}
Cowles MK, Carlin BP. Markov chain Monte Carlo convergence diagnostics: a comparative review. {\it Journal of the American Statistical Association} 1996\string; 91(434)\string: 883--904.

\bibitem{greenwald1998measuring}
Greenwald AG, McGhee DE, Schwartz JL. Measuring individual differences in implicit cognition: the implicit association test.. {\it Journal of personality and social psychology} 1998\string; 74(6)\string: 1464.

\bibitem{nosek2005understanding}
Nosek BA, Greenwald AG, Banaji MR. Understanding and using the Implicit Association Test: II. Method variables and construct validity. {\it Personality and Social Psychology Bulletin} 2005\string; 31(2)\string: 166--180.

\bibitem{henry2002symbolic}
Henry PJ, Sears DO. The symbolic racism 2000 scale. {\it Political psychology} 2002\string; 23(2)\string: 253--283.

\end{thebibliography}
